\documentclass[12pt,a4paper,titlepage]{article}

\usepackage{graphicx}
\usepackage{amsmath}
\usepackage{amsfonts}
\usepackage{amssymb}

\usepackage{color}
\usepackage[numbers]{natbib}
\usepackage[unicode]{hyperref}
\usepackage{times}
\usepackage[T1]{fontenc}

\newcommand{\D}{\,\mathrm{d}}

\begin{document}

\title{KIPT Positron Source Project \\ Conceptual Design}

\author{S.~Perezhogin, E.~Bulyak, A.~Dovbnya, V.~Mitrochenko,\\ A.~Opanasenko, V.~Kushnir}

\date{NSC KIPT, Kharkov, Ukraine 2018}

\maketitle

\pagebreak[4]

\abstract{We present the results of theoretical and numerical studies on the source of slow positrons for NSC KIPT.  The positrons are intended to generate with the electrons of 9, 40 or 90 MeV available at the KIPT electron linacs. The yield of positrons from the conversion target is estimated as well as their spatial--angular characteristics. Optimal parameters of the conversion target for each energy of electrons are estimated. Preliminary design of the positron beam formation system is also presented. Qualitative analytical dependencies of the positron beam parameters at the system exit upon the amplitude and the decrease factor of the magnetic field in the Adiabatic Matching Device (AMD) solenoid have been established. These dependencies have been used for system optimization. Numerical simulations allow to optimize the parameters of AMD for solenoid available in the laboratory. Possible application of the subharmonic RF cavity for reduction of the energy of positrons has also been estimated and validated by the simulations. As it has been shown, this cavity can substantially decrease the positron energy and thus facilitate operation
of the moderator.}

\tableofcontents

\section{Introduction}
Employing the monochromatic positron beams with energy below 1\,MeV provides the unique possibility for research study in the solid materials. The methods of positron annihilation spectroscopy are capable to detect the electron structure of crystals and numerous small-size defects both in solids and porous materials, such as vacancies, vacancy clusters and voids up to size of a cubic nanometer. All these methods have been widely applied in the modern material science, particularly, in the atomic and electronic material science, see \cite{siegel80,singh16}. That is why development of the intensive positron sources based upon the electron linacs is an important yet complicated task.

Commonly, two types of positron sources are employed, radioactive and acce\-le\-ra\-tor-based, \cite{golde12}. The former uses $\beta^+$ active isotopes such as $^{22}$Na or $^{64}$Cu, which directly emit positrons. The sources of this type are compact yet of low intensity, and produce the monoenergetic positrons in MeV range. In the latter -- the electron linac based -- the intensive electron beam is converted into the positron one, with wide spectrum and intensity up to $10^{11}$\,positron per second, by many orders higher then the radioactive source. Due to relativistic effect, the generated positrons are contained within a relatively small solid angle in contrast to the radioactive isotropic emission. Thus efficiency of the accelerator source is much higher. At present time, the positron sources of this kind based on the electron linacs with energy from 10\,MeV to 70\,MeV are employed for study properties of materials by the methods of positron-annihilation spectroscopy \cite{wada13,rourke13,chemerisov09}.

NSC KIPT has a certain experience in construction and employment of a positron source for high-energy physics experiments, see \cite{artemov84}.

\section{Theoretical background}
\subsection{Electron-to-positron conversion}
In the accelerator based method of production, the positrons are generated via two--stage conversion. First, the accelerated electrons produce brem\-s\-strah\-lung radiation while passing through conversion target. Then, these photons produce electron--positron pairs in the strong field of nuclei. The positrons having traversed the conversion target then are collected for further acceleration/deceleration to meet needs of final users.

The photons for positron production must have the energy in access of 1\,MeV: the pair rest energy equals to 1.022\,MeV. With increase of gamma's energy above this threshold positrons production rate mainly increases due to smaller losses of them while traversing the conversion target. For higher energy of the initial electrons, the secondary electrons and positrons have high enough energy to cause electron--positron showers, when these secondary particles produce the gamma-photons, which in turn are converted into the pairs.

It should be noted that besides conversion both the electrons and the positrons suffer from parasitic processes reducing the yield and increasing the emittance of the positron beam \cite{shulga93}. Mainly, these processes are:
\begin{itemize}
\item Bremsstrahlung spectrum declines gradually with energy: Vast part of the photons possess energy below the pair creation threshold. Emission of these photons decreases the energy of initial electrons (so called radiation losses) and increases the angular spread of electrons' trajectories.
\item A define fraction of the energy has been lost due to ionization losses. This process is of importance for the positrons, since probability of annihilation increases with the energy decrease.
\item Elastic scattering (without energy loss) results in increase of angular spread in trajectories.
\end{itemize}

For estimation of the yield of positrons, we separate consideration of the whole process of generation -- beginning from impinging the electrons upon the front (upstream) surface of convertor and finishing at escaping (emitting) positrons from the rear (downstream) surface -- into elemental specific physical processes. Such commonly accepted approach allows to optimize the system aiming at maximum yield of positrons. The processes to be modeled are as follows.
\begin{enumerate}
\item Built up the radiation field in the target bulk by the initial electrons.

\item Creation of the electron--positron pairs in the volume.

\item Motion of the positrons to the rear surface of the target.

\end{enumerate}

The escaped positrons present the positron beam.

For numerical estimations we will employ tantalum and tungsten conversion targets and three energy of the initial electrons: 9\,MeV, 40\,Mev and 90\,MeV, available at existing accelerators of KIPT.

\subsection{Radiation density}
An electron with energy $E_e$ produces in a (thin) target the photons with wide spectrum -- so called bremsstrahlung radiation -- see \cite{koch59}. This spectrum can be approximated by the formula (see \cite{roy68,shulga93}):
\begin{eqnarray} \label{eq:brems}
\frac{\D \sigma}{\D E_p} &=& \frac{4\alpha_{f} (Z r_0)^2}{E_p}\left[ 1+\frac{(E_e-E_p)^2}{E_e^2}-\frac{2}{3}
\frac{(E_e-E_p)}{E_e}\right]\times \nonumber \\
&&\left[
\ln\left(\frac{2 E_e(E_e-E_p)}{m c^2 E_p}\right)-\frac{1}{2}\right]\; ,
\end{eqnarray}
where $E_p$ is the photon energy, $\alpha_{f}$ is the fine structure constant, $r_0$ and $m c^2$ are the classical electron radius and its rest energy, resp., $Z$ is the target material charge number.

The electron energy decreased after a photon emitted, therefore the next photon will be emitted with smaller energy in average. The losses of this kind are referred to as the radiation losses and are dominant in the considered range. The radiation losses are in linear proportion to the electron energy, \cite{nist}. Due to linear dependence of losses, the average electron energy is decreasing exponentially with the decay length equal to the radiation length of target material.

\begin{figure}[htb]
\centering
\includegraphics[width=0.8\textwidth]{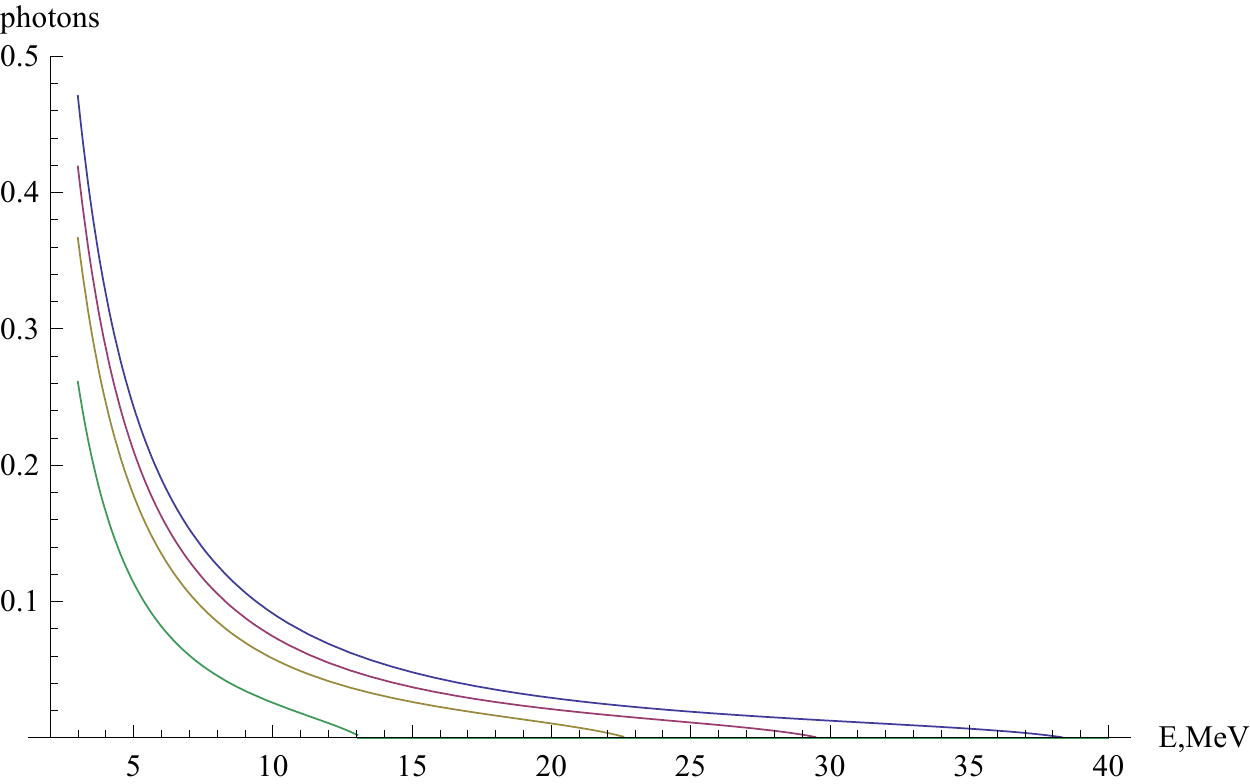}
\caption{Bremsstrahlung spectra in tungsten at 0, 0.25, 0.5, 1 rad. length (from top to bottom) produced by 40\,MeV electrons.\label{fig:sect}}
\end{figure}

The radiation spectra produced by the electrons with the initial energy 40\,MeV at different depth in the tungsten target are presented on figure~\ref{fig:sect}.

In contrast to electrons, which gradually lose their energy but conserve number of them, the high-energy photons while traversing the target lose the number of them (intensity) but preserve the spectrum. Decreasing of the intensity occurs mainly due to conversion of the photons into the electron-positron pairs.

\begin{figure}[htb]
\includegraphics[width=\textwidth]{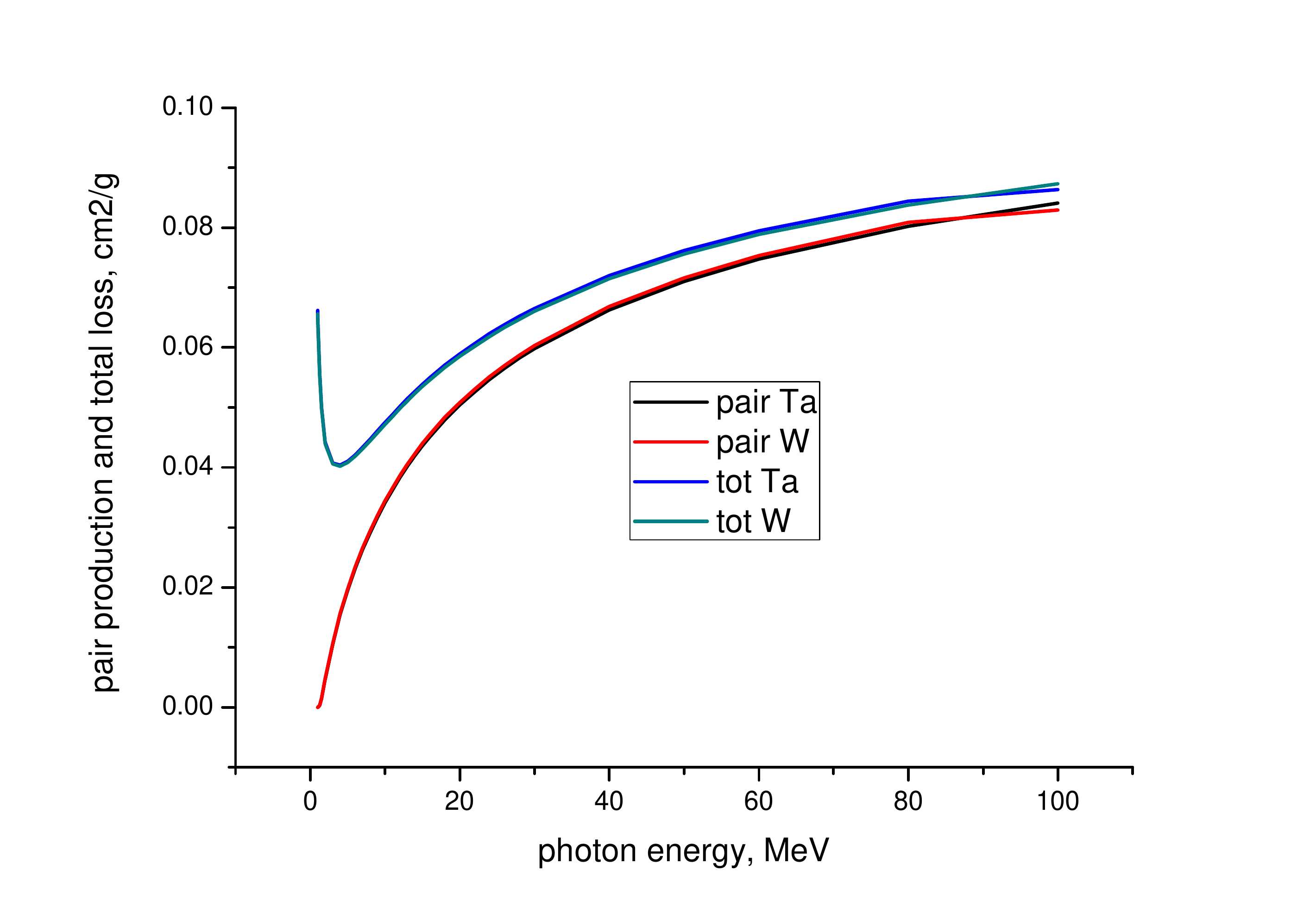}
\caption{Losses of photons and yield of pairs for tantalum and tungsten. \label{fig:photTaW}}
\end{figure}

In figure~\ref{fig:photTaW}, there are present dependencies of the total loss of photons upon their energy and losses due to conversion into the pairs (the data taken from NIST) for tantalum and tungsten (practically the same). As it can be seen, at the energy higher than $\sim 10$\,MeV the photon losses are due to production of the pairs.

\subsection{Positron density}
For not too high energy of the gamma quanta, as in the considering case, every quantum with the energy $E_\mathrm{g} = m_ec^2 \gamma_\mathrm{g} $ produces $\kappa$ electron-positron pairs over one radiation length:
\begin{equation} \label{pairprod}
\kappa(\gamma_\mathrm{g},Z) = \frac{7}{9\ln (183Z^{-1/3})}\left[\ln 2\gamma_\mathrm{g} -\frac{109}{42}- 1.2021(\alpha Z)^2\right]\; ,
\end{equation}
where $Z$ is the charge number of nucleus of the conversion target, $\alpha\approx 1/137$  is the fine structure constant. As it can be seen from this formula, dependence of density of the born pairs upon the energy of quantum is weak (logarithmical).

\begin{figure}[htb]
\includegraphics[width=\textwidth]{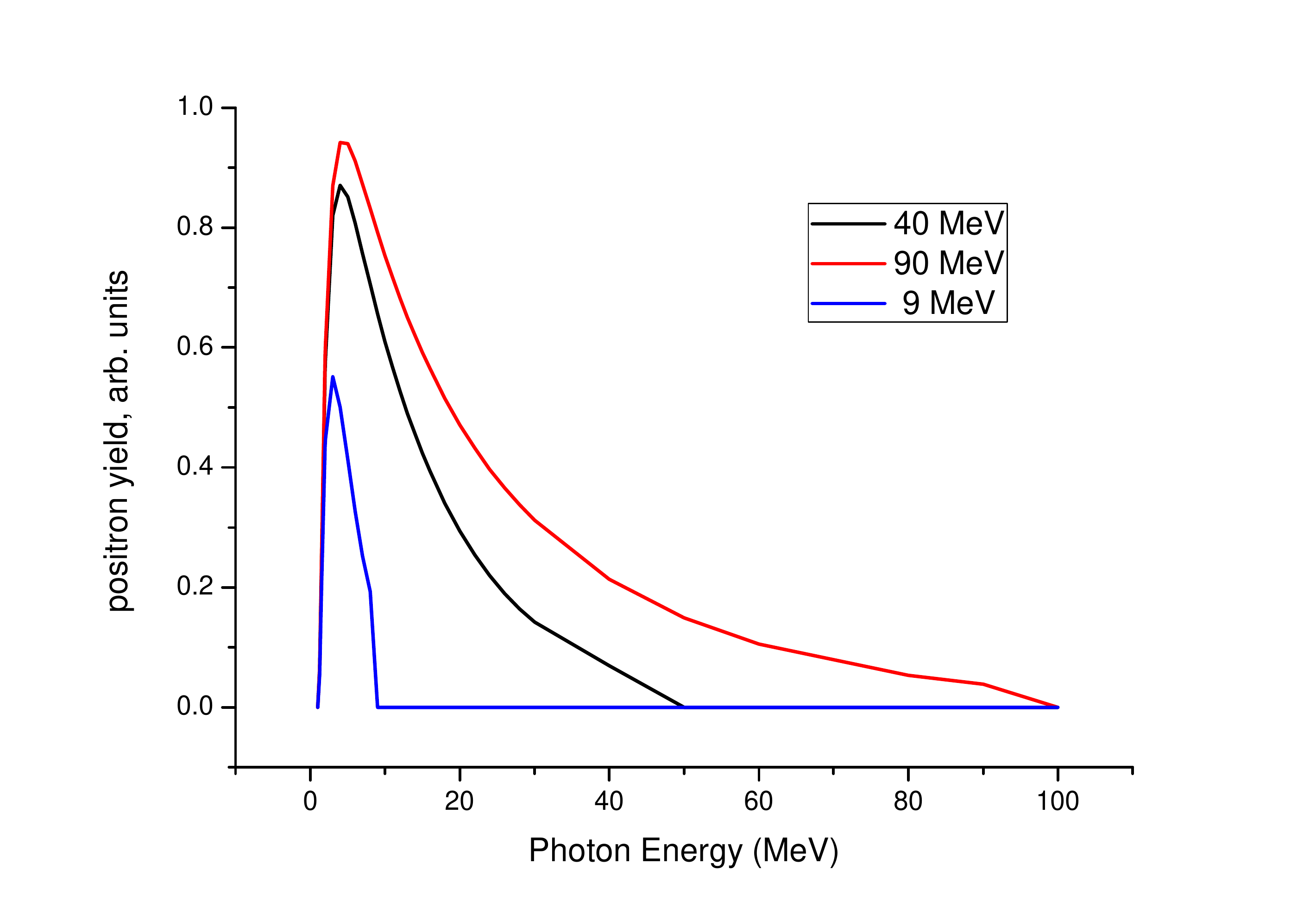}
\caption{Yield of the pairs born by bremsstrahlung, arb. units.\label{fig:yeld94090}}
\end{figure}

Yield of the positron-electron pairs in tungsten converter via the energy of gammas computed with the code XCOM of NIST, is presented in figure~\ref{fig:photTaW}. The energy of gammas to produce the pairs should be higher than 1.25\,MeV. Convolving the bremsstrahlung spectrum \eqref{eq:brems} with probability of the pair production \eqref{pairprod}, we may deduce the spectrum of total energy of pairs -- electron+positron -- as function of the  energy of initial electron. Figure~\ref{fig:yeld94090} presents distributions of the energy of pairs for the energy of primer electrons 9, 40 and 90\,MeV.

As it is seen from the curves in this figure, maximum number of pairs produced at the energy far below the initial energy of the electrons: within the range from 5\,MeV to 10\,Mev for the electron's energy from 9 up to 90\,MeV. It is important to emphasize that total number of the pairs -- area encircled by the corresponding curve -- sufficiently increases with increase of the electron's energy.

The positrons are born in the volume of the target bulk, their spectrum and  density are determined by the initial energy of the impinging electrons and distance from the front face of the target. As it follows from the considered above theory, each stage min the process of positron production, $\mathrm{e}^{-}\to\gamma\to \mathrm{e}^{+}+\mathrm{e}^{-}$ increases the spectral width of the corresponding ensemble and shifts it toward lower energy. As a result, the maximum number of positrons are produced with a smaller energy than that of the prime electrons.

\subsection{Positron stream from the target}
Born in the target bulk positrons are moving toward the rear face of the conversion target, their energy degrades along the pass. Unlike electrons, the number of positrons is decreased because of annihilation with electrons. Beyond the annihilation, interactions of the positrons with matter are practically the same as of the electrons: the radiation losses dominate in degradation of the energy. The ionization losses that are dominant at the energy below a few MeVs, are equal to that of the electron's with precision better than 10\,\%, see \cite{roy68}.

The annihilation cross-section has the  inverse dependence on the energy. The positrons losses due to two-photon annihilation (dominant mechanism) read:
\[
\frac{\partial N}{N\partial t} = - \frac{\pi}{4\alpha}\, \frac{(\ln 2\gamma_+ -1)}{\gamma_+}\, \frac{A}{Z\Lambda(Z)}\; ,
\]
where $t$ is the pass length in rad.length, $\gamma_+$ the energy (Lorentz-factor) of the positron, $\Lambda(Z)=\ln 183/Z6[1/3]$ the Coulomb logarithm, $A$ the atomic number of the target nuclei, $Z$ the charge number.

\subsection{Optimal thickness of the target}
As it can be seen from the above, the electrons lose their energy sufficiently faster then the gamma quanta generated by electrons. (The difference in degradation of the electrons moving through matter and the photons is that the electrons lose their energy preserving the number, while the photons preserve their energy decreasing the number of them.)

In accordance with this mechanism, the density of gammas, and thus the density of the produced electron-positron pairs, will increase with the target thickness, reaches a certain maximum. Then it will be exponentially degraded when the electron's energy decreases to the limit when the generated gammas will no be able to produce positrons -- the threshold energy of electrons is about 2\dots3\,MeV.

When the born positrons move toward the rear face of the conversion target, their energy and number are decreased.

When the target is thin, a small number of gammas is produced and thus small number of the positrons. On the other hand, at large thickness of the electrons' kinetic energy transfers into gammas and then to positrons. But the latter are annihilated in the target body. As it follows from the reasoning, there exists an optimal target thickness that produces maximum number of positrons.

\begin{figure}[htb]
\includegraphics[width=\textwidth]{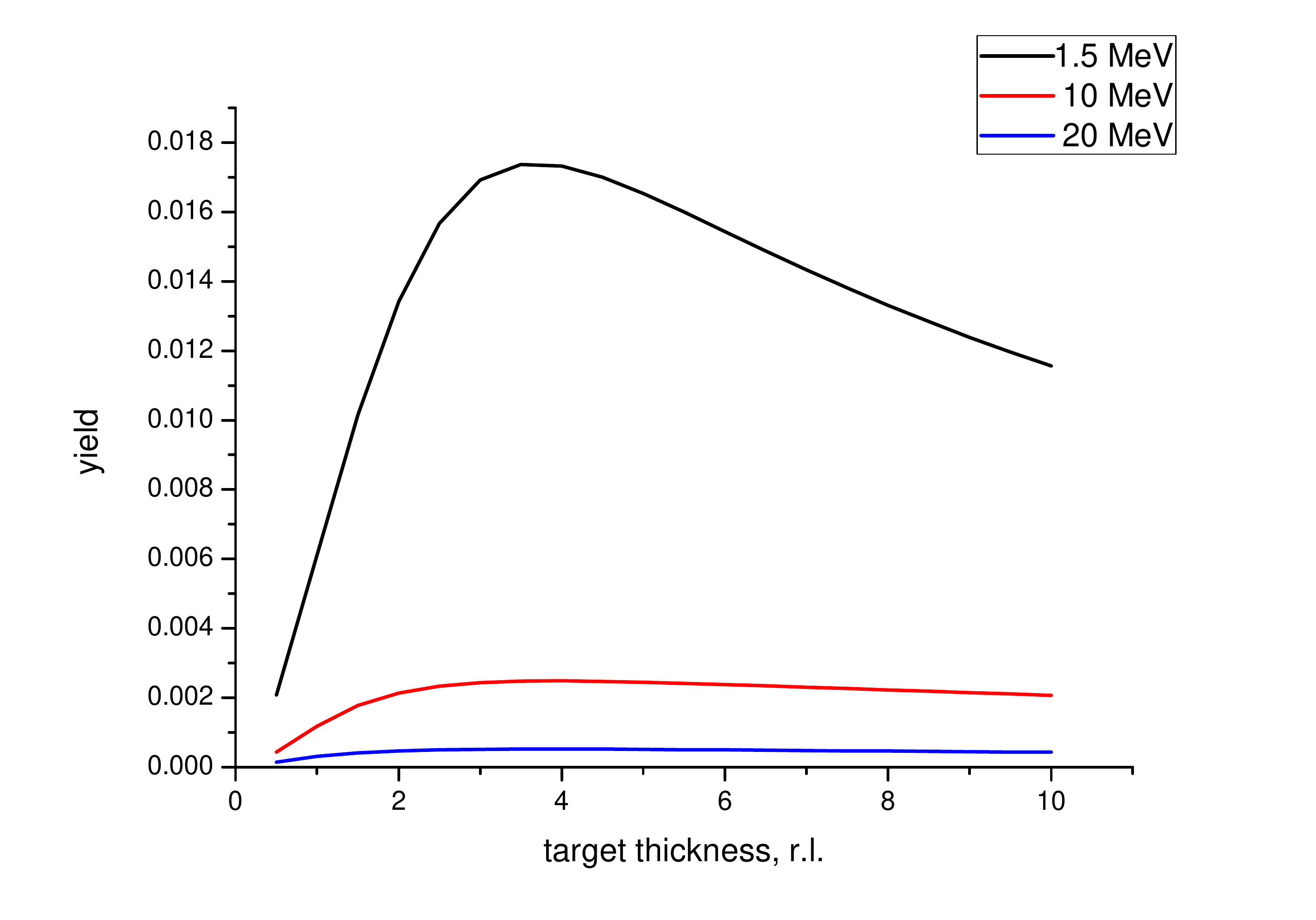}
\caption{Yield of 1.5, 5 and 10\,MeV positrons from the tantalum target. \label{fig:yield40L}}
\end{figure}

In figure \ref{fig:yield40L} there are presented the dependencies of yield of the positrons with the energies of 1.5, 5 and 10\,MeV from tantalum target on it's thickness (in rad.length) for the 40\,MeV electron beam.

As it can be seen, the yield is weakly dependent on the positrons' energy and reaches  maximum at the thickness 2\dots 4 r.l.

As a result of study on yield of the positrons from the target via the energy of the initial electrons, it was established that number of the gammas with the `convertible' energy logarithmically slow increases with the energy of electrons. In turn, cross-section of the pair production has the logarithm dependence on the energy of quanta. Thus, dependence of the number of produced positrons upon the electrons' energy is rather weak.

Nevertheless, efficiency of positrons production significantly increases with the energy of electrons due to the fact that the more energetic electrons produce larger fraction  of the energetic gammas, that are converted in the more energetic positrons. The loss rate of high energy positrons is smaller than that of the slow ones.

\section{Computing on the positron source}
Theoretical considerations yield a general picture of the process -- ``dependencies''. To obtain the specific numbers we used a CERN-originated code, GEANT4 \cite{geant4}. For verification of the model, we used the scheme adjusted for registration of the well-known bremsstrahlung radiation.

\begin{figure}[htb]
\includegraphics[width=\textwidth]{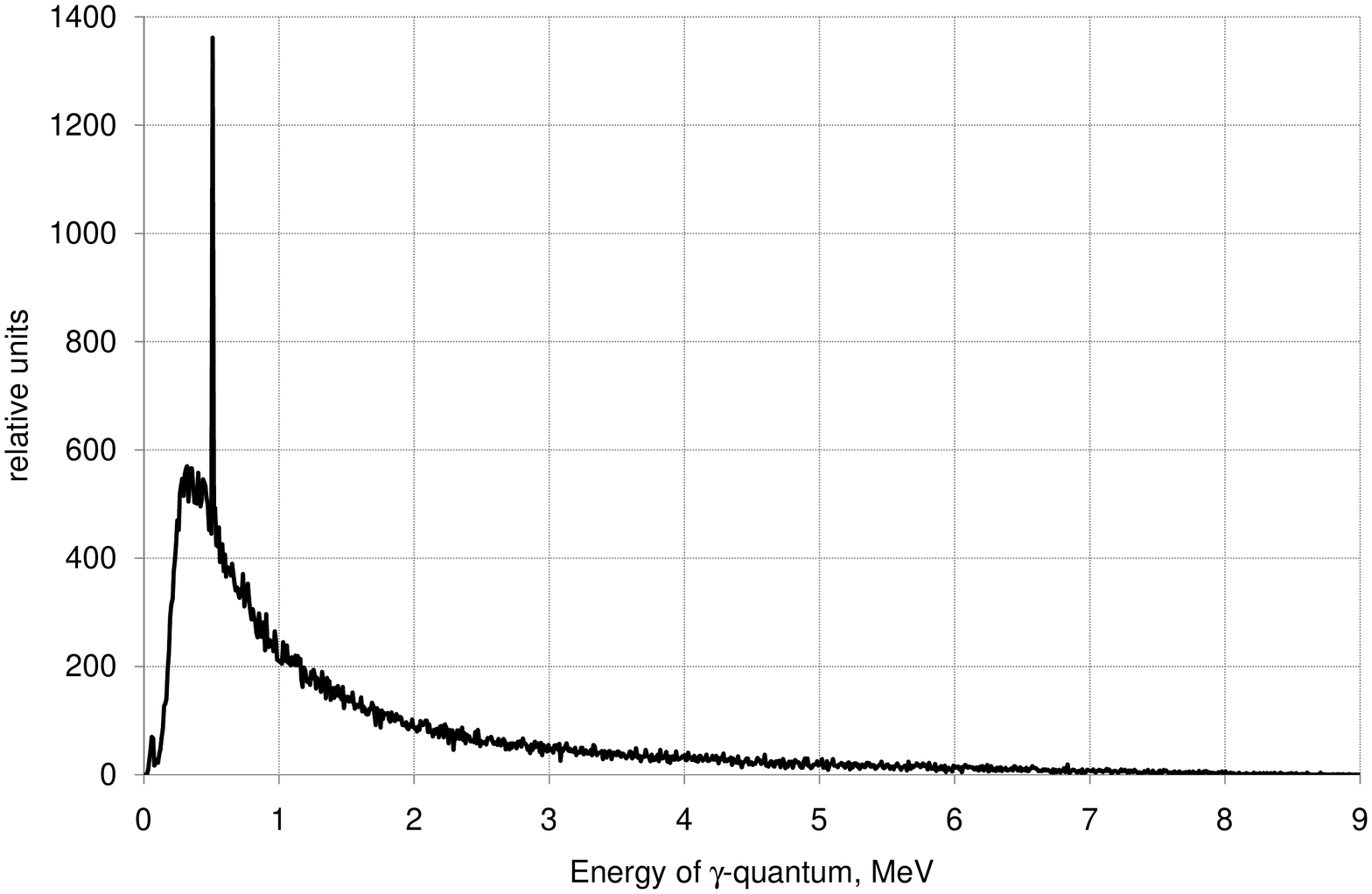}
\caption{Gamma-spectrum from tantalum target of 2\,mm thickness, irradiated with 9\,MeV electrons. \label{fig:7}}
\end{figure}

A test computation shows that the model is viable and meets the requirements: We obtain the specific spectrum of the bremsstrahlung radiation superimposed by the spectrum of the two-photon annihilation of positrons, see figure \ref{fig:7}.

A major goal of simulations was to obtain the spectra of positron energy emitted from the conversion target together with the total yield $N_{p}$ as ratio to the total number of income electrons $N_e$ as function of the target thickness.

\begin{figure}[htb]
\centering
\includegraphics[width=\textwidth]{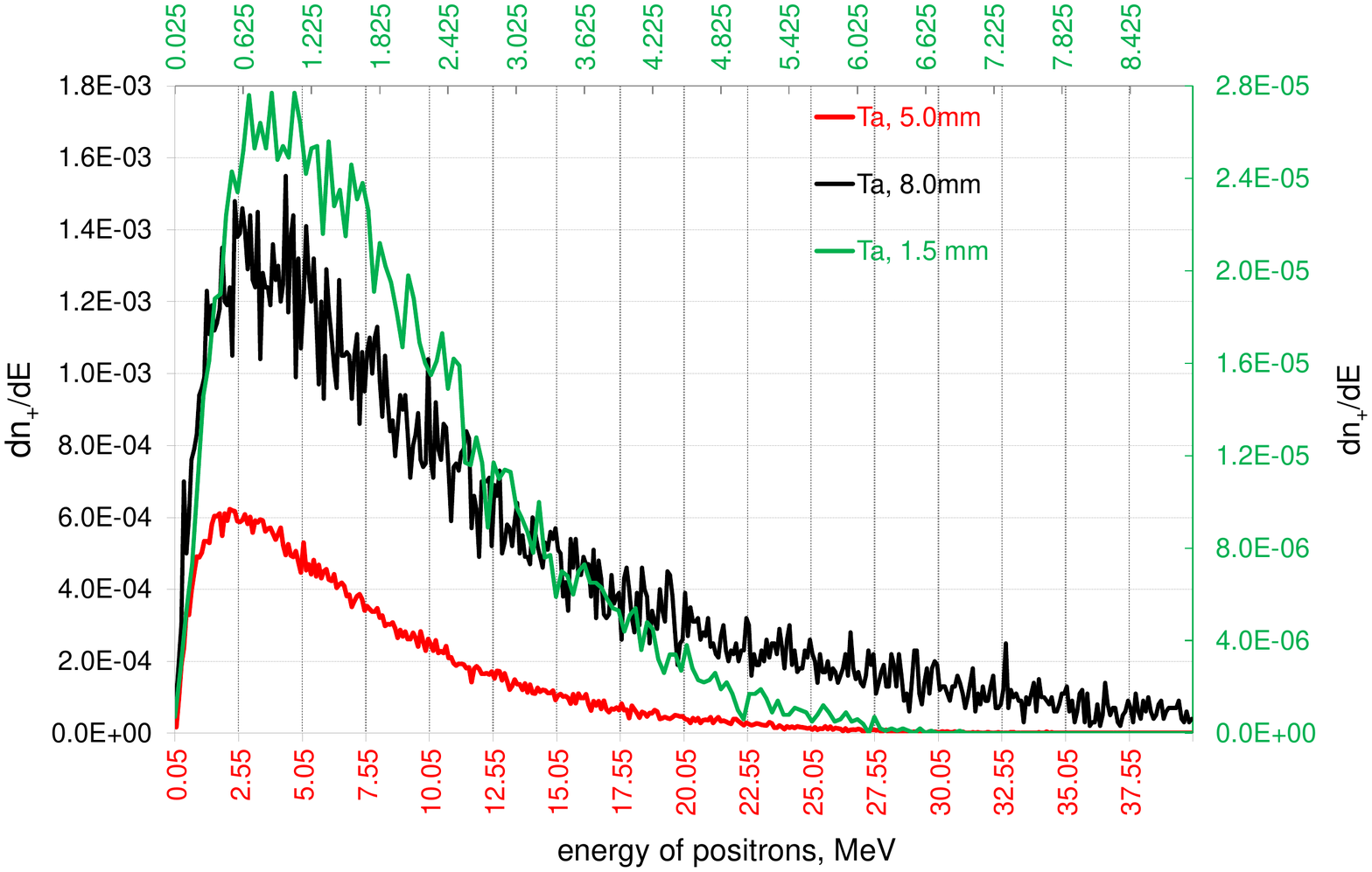}
\caption{Spectra of positrons at target's exit: green line for 9\,MeV electron, red for 40\,MeV and black for 90\,MeV. \label{fig:8}}
\end{figure}

\begin{figure}[htb]
\centering
\includegraphics[width=0.8\textwidth,trim=70 50 0 200]{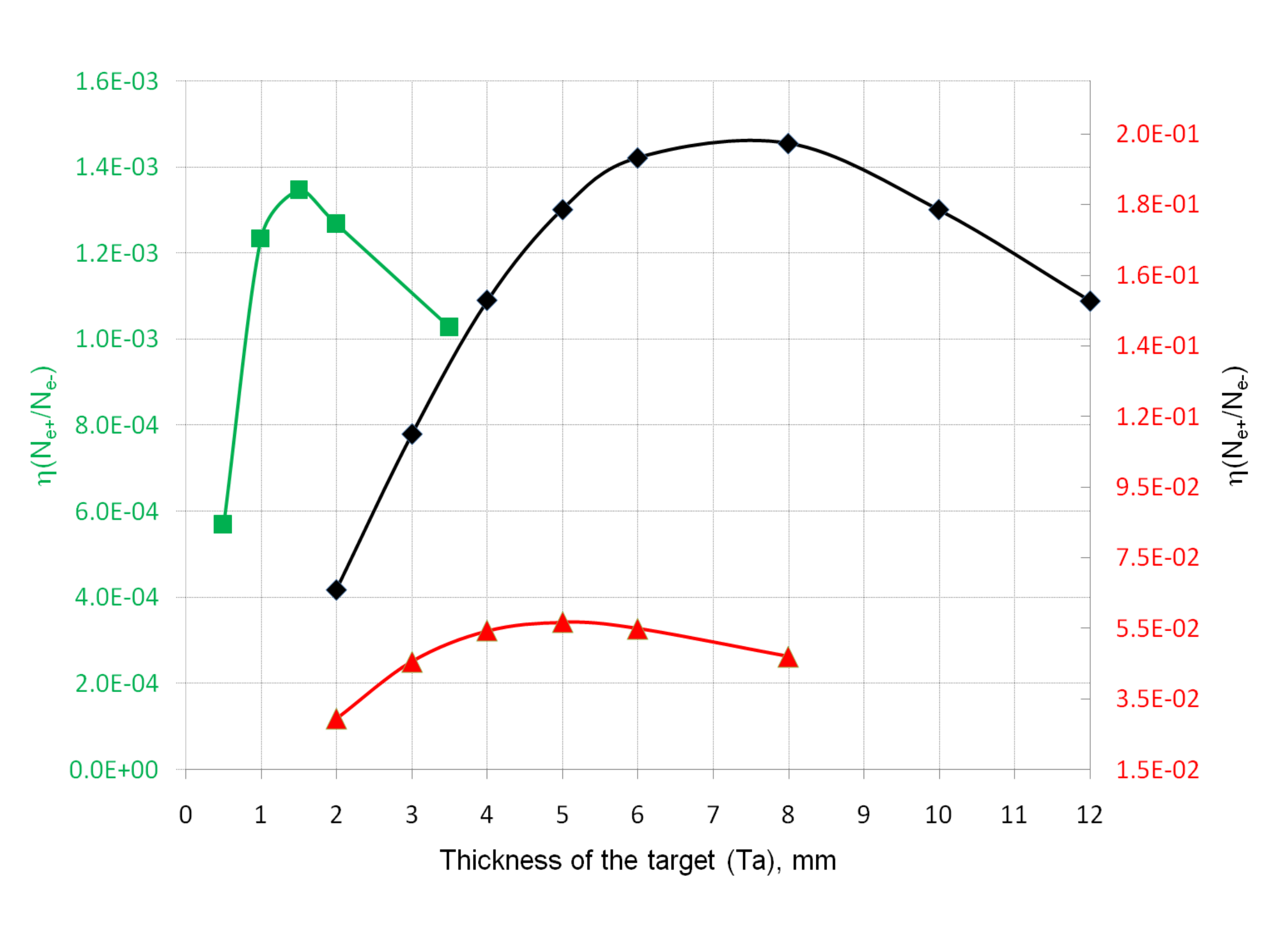}
\caption{Efficiency of positron production via the target thickness for electrons of 9\,MeV (gren), 40\,MeV (red) and 90\,MeV (black). \label{fig:9}}
\end{figure}

Simulations for the electron energies of 9\,MeV, 40\,MeV and 90\,MeV result in determination of the optimal thicknesses of the targets. The spectra of positrons per initial electron for the simulated energies are presented on figure~\ref{fig:8} for the optimal converters. Efficiency of positron production (ratio of number of positrons per impinging electron) is presented in figure~\ref{fig:9}.

The results of simulation proved qualitative dependence of the yield of positrons on the energy of electrons and the target thickness. The quantitative results were sufficiently clarified.

The simulations display the presence of large amount of isotropically distributed annihilation quanta. Isotropic distribution indicates that the positrons have been retarded down to nonrelativistic energy before annihilation. A detector for the photons from annihilation of positrons would be a good monitor of the positron source.

Main results of the simulation are presented in table~\ref{tab:1}.

\begin{table}[hbtp]
\caption{The results from simulation.}
\label{tab:1} \centering
\vspace{6pt}
\begin{tabular}{|c|c|c|c|c|}
\hline
$E$ &$N_e$ & $N_p$ & $\eta=N_p/N_e $  & $d$ \\[1ex]
MeV &  &  &   & mm/r.l. \\[1ex]
\hline
9 &  $10^{7}$ & 13466& $(1.350\pm 0.010)\times10^{-3}$ & 1.5 / 0.4 \\[1ex]
40 &  $10^{6}$ & 56772 & $(5.677\pm 0.024)\times10^{-2}$ & 5.0 / 1.3\\[1ex]
90 &  $10^{5}$ & 19235 & $(1.924\pm 0.013)\times10^{-1}$ & 7.0 / 1.8\\[1ex]
\hline
\end{tabular}
\end{table}

As it follows from the results of simulations, the total yield of  positrons from the tantalum conversion target of optimal thickness is dramatically dependent upon the energy of initial electrons: increase in the energy from 9\,MeV up to 90\,MeV will increase the yield by more than two orders of magnitude. This enhancement in the yield requires a more thick conversion target.

Estimates for the positrons yield from the linacs existed in NSC KIPT: LUE-10 ($E=9$\,MeV, the electrons' average current $270\,\mu$A) and LUE-40 \cite{dovbnya14} with the energy variate in the range 40--90\,MeV at the average current $5\,\mu$A show that expected positron yield may be as high as $2.3\times 10^{12}$\,pos/s (LUE-10) and $(1.8\dots 5.9)\times 10^{12}$\,pos/s (LUE-40), see table\,\ref{tab:1}.

A classical setup of installation to produce a monochromatic positron beam comprises the positron source -- the isotopes or the linac with an electron-to-positron converter, and a moderator attached downstream. The moderator consists of the material  with negative positron affine (tungsten or inert gases in solid state). As a result of bombardment, the positrons are thermalize, a fraction of them with energy of a few eV diffuses back on the surface. These `slow' positrons are accelerated in constant electric field and are transported to the object under examination. Since the positron beam is of big angular spread, the moderator should be placed as close to the source as possible.

The moderator efficiency (ratio of the `slow' positrons fraction to the total number of them) substantially reduces with increase of the energy of initial positrons and reaches $\sim 10^{-5}$ at the energy of a few MeV, \cite{rourke11}. One of effective methods to reduce the energy of the income to the moderator positrons was proposed in ANL, \cite{long07}. The method consists in set up a RF cavity in front of the moderator. The cavity slows down the positrons. Naturally, it requires a system to transport the positrons from the converter to the cavity

\section{Design of the positron transport system}
The positron beam that produced in the conversion target possesses specific properties:  large spread of energies (wide spectrum), large angular spread and a relatively small radius, order of a millimeter. Such set of the parameters, especially the wide angular spread, does not allow directly to accelerate/decelerate the beam and to transfer it to the experimental site. The positron beam should be properly formatted and refined from the electrons and the residual gammas. In order to match the beam emittance to the acceptance of the transport line, the Adiabatic Matching Device (AMD) is employed most commonly. While passing along AMD, the shape of emittance is transforming from large angular spread and small radius into a relatively small angular spread and a large radius.

\subsection{Theoretical estimations}

For preliminary estimation we use so-called paraxial approximation,
which is valid for relatively small angles between a positron trajectory  and the axis of the system. Within this approximation, transversal motion of a positron can be considered as nonrelativistic with the transversal mass $m = \gamma m_0$ ($\gamma $ is the Lorentz-factor of a positron).

In the cylindrical framework suitable for a problem with the axial symmetry we can derive the nonrelativistic Hamiltonian. Let us start from the Lagrangian
\begin{equation} \label{eq:lagr}
\mathcal{L}=\frac{mv^2}{2}+e\vec{A}\cdot\vec{v}\; ,
\end{equation}
where $\vec{A}=A_\theta(r,z)$ is the (vector) potential of the magnetic field, which can be described by the single axial component.

Substituting the velocity components into \eqref{eq:lagr},
\begin{align}
v_r &= \dot{r}\; ; & v_z &=\dot{z}\; ; & v_\theta = r\dot{\theta}\; ,\nonumber
\intertext{we get the canonical momenta}
p_r &= \frac{\partial \mathcal{L}}{\partial\dot{r}}=m\dot{r}\; ; &
p_z &= \frac{\partial \mathcal{L}}{\partial\dot{z}}=m\dot{r}\; ; & p_\theta &= \frac{\partial \mathcal{L}}{\partial\dot{\theta}} = m r^2 \dot{\theta}\; . \label{eq:canonp}
\end{align}

From the relation
\[
\mathcal{H} =\sum_i \dot{q}_ip_i - \mathcal{L}
\]
accounting for \eqref{eq:lagr} and\eqref{eq:canonp}, we deduce an expression for the Hamiltonian in the axially symmetric field in the cylindrical coordinate frame:
\begin{equation} \label{eq:hamilt}
\mathcal{H}=\frac{p_r^2+p_z^2}{2m}+\frac{1}{2m}\left(\frac{p_\theta}{r}-eA\right)^2\; ,
\end{equation}
with $A\equiv A_\theta$.

From \eqref{eq:hamilt}, there come the canonical equations describing the particle trajectory. The equations for the coordinates are
\begin{subequations} \label{eq:canonco}
\begin{align}
\dot{r} &= \frac{\partial\mathcal{H}}{\partial p_r}=\frac{p_r}{m}\; ;  \\
\dot{z} &= \frac{\partial\mathcal{H}}{\partial p_z}=\frac{p_z}{m}\; ; \\
\dot{\theta} &= \frac{\partial\mathcal{H}}{\partial p_\theta} = \frac{1}{mr}\left(\frac{p_\theta}{r}-eA\right)\; . 
\end{align}
\end{subequations}
for the momenta:
\begin{subequations} \label{eq:canonmo}
\begin{align}
 \dot{p}_r &= -\frac{\partial \mathcal{H}}{\partial r} = \frac{1}{m} \left(\frac{p_\theta}{r}-eA\right)\left(\frac{p_\theta}{r^2}+e\frac{\partial A}{\partial r}\right)\; ; \\
\dot{p}_z &= -\frac{\partial \mathcal{H}}{\partial\dot{z}} = \frac{1}{m} \left(\frac{p_\theta}{r}-eA\right)\left(e\frac{\partial A}{\partial z}\right)\; ; \\
\dot{p}_\theta &= -\frac{\partial \mathcal{H}}{\partial \theta } = 0\; , \label{eq:canontheta}
\end{align}
\end{subequations}
and for the temporal evolution of the Hamiltonian
\begin{equation} \label{eq:hamt}
\frac{\D \mathcal{H}}{\D t} = \frac{\partial \mathcal{H}}{\partial t} = 0\;.
\end{equation}

The two integrals of motion come from \eqref{eq:hamt}  and \eqref{eq:canontheta}: the Hamiltonian itself equal to the total energy of a particle, which is independent explicitly   of time
\begin{subequations} \label{eq:const}
\begin{align}
\mathcal{H}&=\mathrm{const}\; , \label{eq:consth}
\intertext{
and conservation law for the angular momentum:}
\dot{p}_\theta &=  0\;\to\; p_\theta = mr^2\dot{\theta} + e A r = \mathrm{const}\; .\label{eq:constpt}
\end{align}
\end{subequations}

\subsection{Constant--field solenoid }
The simplest case is the constant magnetic field -- the radially uniform solenoidal field
\[
B_r=B_\theta=0\;,\qquad B_z = B\; .
\]

Assuming the only axial component of the vector potential being nonzero, from the expression
\[ \nabla \vec{A} = \vec{B} \]
we can find
\begin{equation} \label{eq:pota1}
\frac{1}{r}\frac{\partial (r A_\theta )}{\partial r} = B\quad\to \quad A_\theta = \frac{Br}{2}\; ,
\end{equation}
where the constant of integration is set up to zero.

As it seen from \eqref{eq:pota1}, the equilines of the magnetic potential (force lines) are parallel to the solenoid axis. Considering the initial conditions (at $t=0$) -- $\theta = \dot{\theta} = 0$, $r = 0$, $\dot{r} = \beta_0 c $ (where $\beta_0\equiv \beta_\perp (t=0)$) -- we get the projection of the trajectory onto the transverse plane:
\begin{equation} \label{eq:rtraj}
r = \frac{\beta_0 c}{\omega }\left(1 - \cos \omega t\right)\; ,
\end{equation}
where $\omega = \omega_\text{cycl}/2\gamma $, $\omega_\text{cycl} = e B /m = 1.7588047\times 10^{11}\times B[\text{Tl}]\,\text{s}^{-1}$.
The positron is oscillating around the axis at the frequency equal to half of the cyclotron frequency. The maximum deviation from the axis is:
\begin{equation} \label{eq:rmax}
r_\text{max} = \frac{2 \beta_0 c}{\omega }=\frac{4\gamma c m \beta_0 }{e B }
\approx \frac{4\gamma c \sqrt{1-\gamma^{-2}}\sin\phi_0}{1.7588\times 10^{11} B[\text{Tl}] }
\; ,
\end{equation}
where $\phi $ is the initial angle with respect to the axis.

In the cylindrical coordinate frame with the center in  $R=r_\text{max}/2, \Theta = \pi/2$, the positron moves with constant speed along a circle of the radius $r_*=r_\text{max}/2$, with angular frequency $\omega_*=2 \omega = \omega_\text{cycl}/\gamma $. Spatial period of such helical motion is
\begin{equation} \label{eq:Lz}
L_z = \frac{2\pi \gamma c}{\omega_\text{cycl}}\cos\phi_0 \sqrt{1-\gamma^{-2}}=\frac{\pi}{2}r_\text{max}\cot\phi_0 \; .
\end{equation}

The positron trajectory represent Larmour ring of radius $r_\text{max}/2$ rotating with frequency $\omega_\text{cycl}$. The center of rotation is displaced by $r_\text{max}/2$ from the axis.

A so-called ``single-particle emittance'' (the envelope of transverse projection)  determined as
\[
\epsilon_{env} =r_\text{max}\frac{\beta_0}{\beta_\parallel} = \frac{4c \gamma \sqrt{1-\gamma^{-2}}}{\kappa B}\,\frac{\sin ^2\phi_0}{\cos\phi_0}
\]
is inversely proportional to the field amplitude $B$. So is the envisaged emittance of the positron beam.

\subsection{Tapered field}
As usual, see e.g. \cite{chehab83,chehab94}, the field decreasing along the solenoid axis is formed according to a dependence
\[
B_z(r=0) = \frac{B_0}{1+\alpha z}\; ,
\]
where $\alpha $ is the slope factor.

This field strength may be represented by a potential
\begin{equation} \label{eq:potz}
A_\theta(r,z) = \frac{B_0 r}{2(1+\alpha z)}\; .
\end{equation}
Axial decrease gives rise to the radial component:
\[
B_r(r,z) =\frac{B_0 \alpha r}{2(1+\alpha z)^2}\; .
\]

For the considering system with a large longitudinal size $Z$ as compared to the transverse one, $r_\text{max} /Z\ll 1$, the introduced above factor $\alpha$ is small: the particle performs many turns while traversing along the system.

\subsection{Adiabatic analysis of AMD}
A positron trajectory in the magnetic field of AMD described by the system of equations \eqref{eq:hamilt},\eqref{eq:canonco},\eqref{eq:canonmo} with the potential \eqref{eq:potz} is rather complicated. A sufficient simplification can be achieved under assumption of small transversal particles' velocity together with a small longitudinal gradient of the field.

These assumptions allow one to average the trajectory over the cyclotron frequency, then to consider dynamics of the `Larmor rings'. A small gradient of the field allows to employ the adiabatic theorem (for the charged particles dynamics in the magnetic field this theorem is referred to as the Bush theorem, see \cite{lawson}). Tapering of the longitudinal component of the field gives rise to the radial gradient and the drift of the Larmor rings in the direction perpendicular both to the gradient and the field strength lines -- so called angular drift: the centers of the rings remain at the initial radius with slow azimuth turning.

The adiabatic approach is applicable if the particle turns many times while traversing the system, the frequency of such transversal oscillations may be considered as the slow function of the time (or the longitudinal coordinate). In this case the most slowly changing parameter -- the adiabatic invariant \cite{bakay81e} -- 
is not the `transversal energy'
\[
\mathcal{H}_\perp =\frac{p_r^2}{2m}+\frac{1}{2m}\left(eA\right)^2=\frac{\gamma m \beta_0^2 c^2}{2}\; ,
\]
but its ratio to the frequency:
\begin{equation} \label{eq:adia}
\mathcal{H}_\perp / \omega(z) \approx \mathrm{const}\; .
\end{equation}

From \eqref{eq:adia}, the main principle of the beam transform directly follows by AMD: due to $D$ times decrease of the field, $D = 1+\alpha Z$, the Larmor radius is increased by $\sqrt{D}$ times, and the transverse component of the velocity decreased also by $\sqrt{D}$ times. `The single-particle emittance' is almost not changed:
\begin{equation} \label{eq:emit1}
\epsilon(Z) = r_*(1 + \sqrt{D}) \times \beta_0 c /\sqrt{D} = \epsilon(Z) \frac{1 + \sqrt{D}}{\sqrt{D}}\; .
\end{equation}
Here we assumed that the center of Larmor ring remains at the initial radius.

\begin{figure}[htb]
\centering
\includegraphics[width=0.5\textwidth]{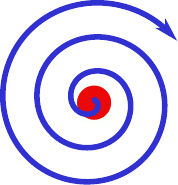}
 \caption{AMD scheme. The positrons at the target exit are red, the transverse projection of a trajectory is in blue.}
   \label{fig:amd}
 \end{figure}

Mitigation of the `transverse energy' by $D$ times is compensated by increasing its longitudinal component -- the positrons trajectories straighten as is schematically displayed in figure~\ref{fig:amd}.

Rough numbers on the positrons beam transformations with AMD are as follows. A positron with the energy 2\,MeV emitted at the angle of 0.1\,rad to the axis of AMD with the field at entrance 1\,Tl, maximally deviates by 2.6\,mm ($r_* = 1.3$\,mm). The initial longitudinal advance per period of transverse rotations will be around  4\,cm. So, at the length of the solenoid about 1\,m (usual AMD length is about $40\dots 80$\,cm) and the  field reduction equal 10 -- the field strength at the exit is 0.1\,Tl -- the final beam radius will be about 5.2\,mm, with 3 times reduction of the angular spread, up to  0.033\,rad.

\section{Numerical simulations}
To elucidate the theoretical estimations, we performed numerical simulations. The numerical simulations of the processes of generation the positron beam as well as its dynamics have been carried out for a system consisting of the 9\,MeV electron linac, the electrons-to-positrons converter, the transport line and the RF cavity for reduction the positrons energy. The packages GEANT \cite{geant4} and PARMELA \cite{parmela} were employed for these purposes.

The simulations was aimed at clarification the production and transport of the low-energy positrons. In addition, we considered moderation (slowing down) of such positrons, implying that the yield of positrons from a moderator is inversely proportional to their energy, \cite{rourke}.

The PARMELA code allows to simulate the spatial dynamics of charged particles in the magnetic field together with RF cavities. PARMELA was applied twice: first for simulation of the electron beam dynamics in the linac and second -- for simulation of the positron beam dynamics downstream after the conversion target. The conversion process was simulated by the code GEANT.

\begin{figure}[htb]
\centering
\includegraphics[width=\textwidth]{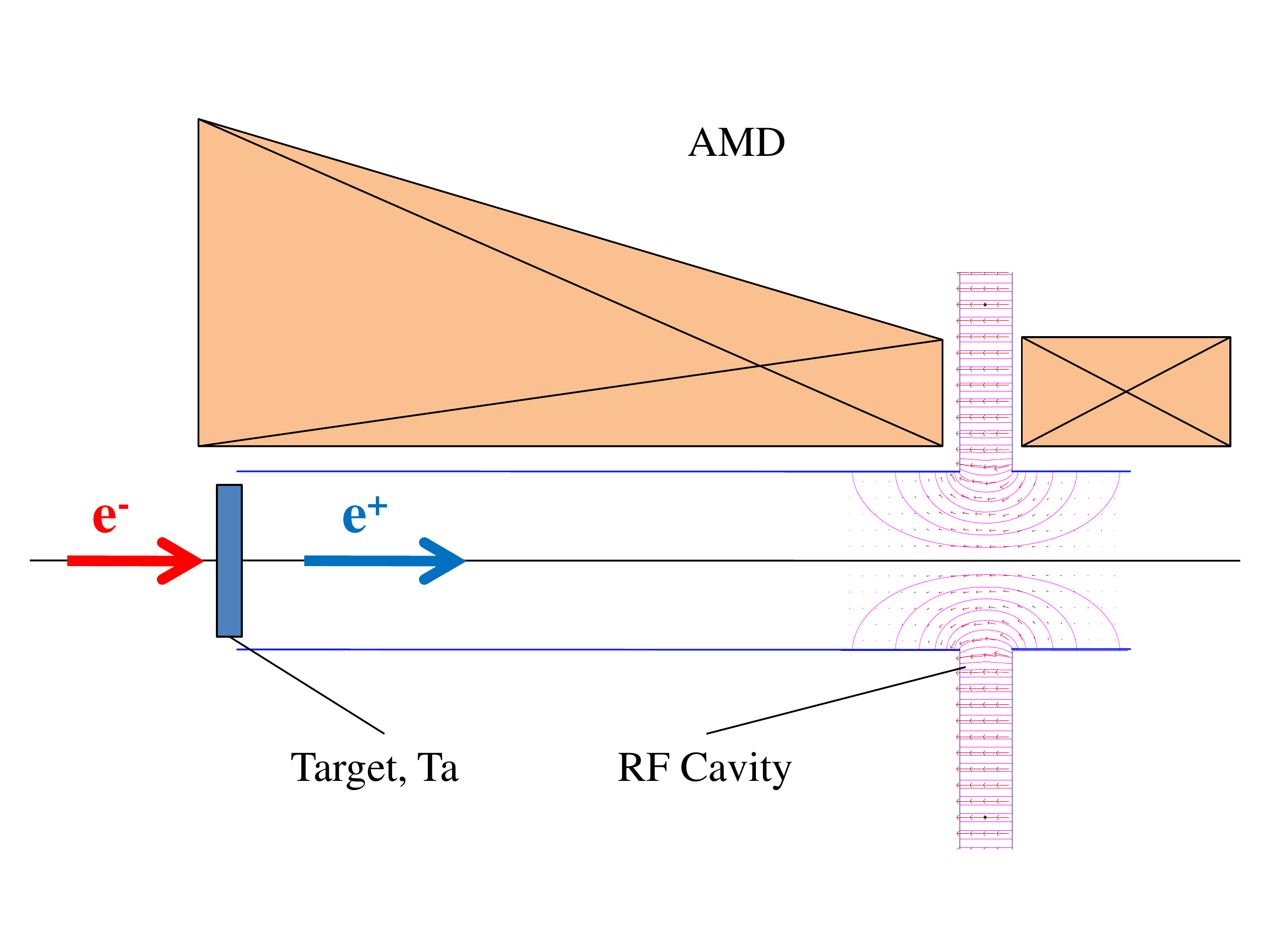}
\caption{Low-energy positrons transport system} \label{fig:10}
\end{figure}

Figure \ref{fig:10} presents a layout of the transport system used in simulations. Between the conversion target and RF cavity, there inserted is AMD (intended mainly to rotate the transverse phase ellipse by $\pi /2$) that allows to reduce the transverse momentum of positrons at the AMD exit.

The AMD magnetic field strength decreased from 1\,Tl to 0.08\,Tl over 75\,cm length. The field decreases in accordance with \eqref{eq:potz}, the parameter $\alpha =0.13$.

\begin{figure}[htb]
\includegraphics[width=\textwidth]{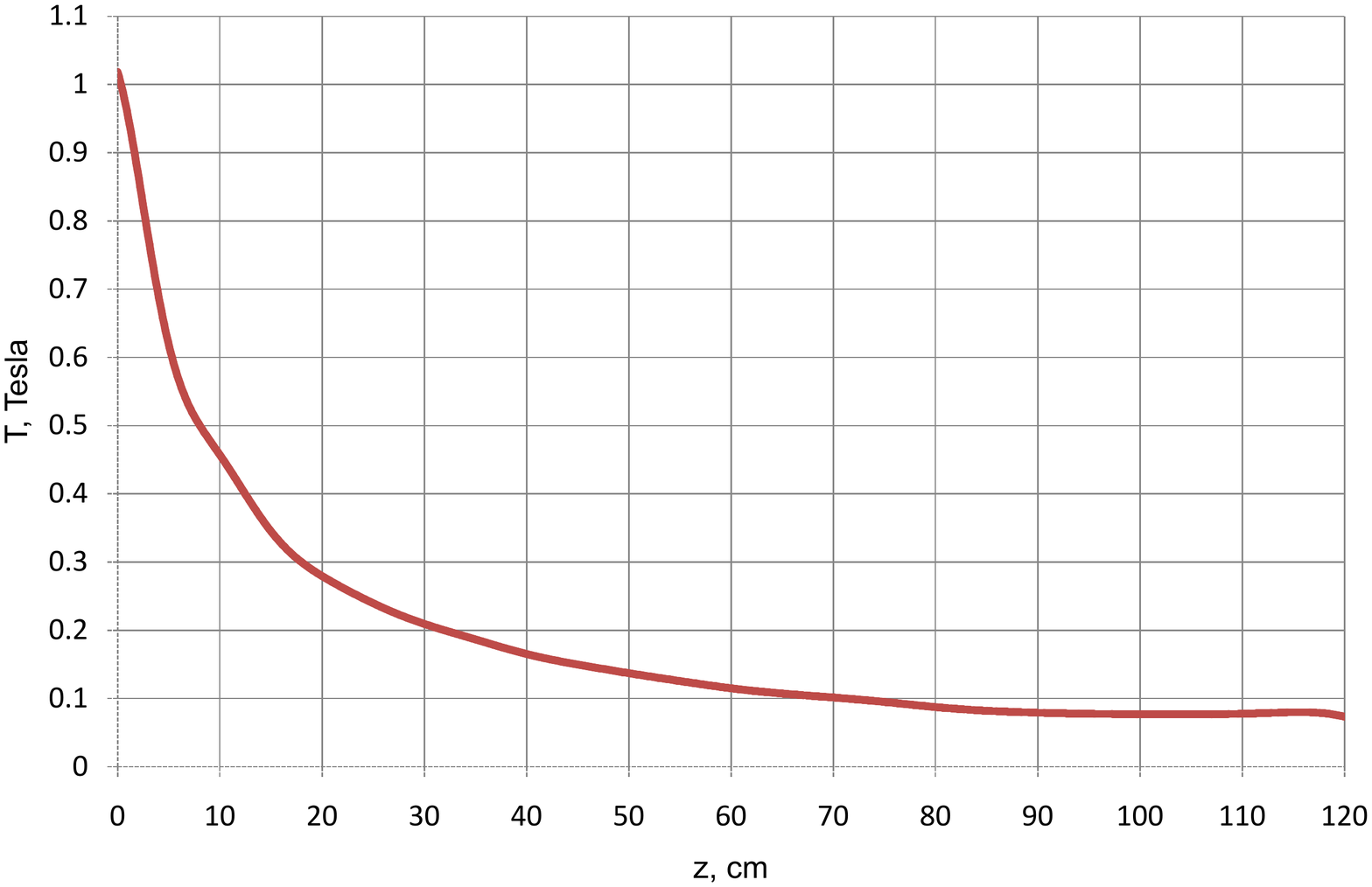}
\caption{AMD field strength along the axial coordinate.  \label{fig:11}}
\end{figure}

Figure \ref{fig:11} represents the on-axis field strength, begin from the target rear plane downstream to the rear end of the RF cavity. Both the AMD field index $\alpha $ and the length were optimized to obtain minimal r.m.s. angular spread of the positrons trajectories.

The RF cavity adjoined to the AMD rear end is intended for slowing down the positron beam. The radius of internal hole of the cavity was 5\,cm in the simulations, the magnetic field strength along the drift space and the cavity was setup to 0.076\,Tl. The RF cavity design (see figure \ref{fig:10}) was performed with the SUPERFISH code \cite{superfish}. The RF basic frequency is 114\,MHz that equal to 25th subharmonics of the accelerator frequency, $f_\text{linac}=2856\,\text{MHz}$.

\begin{figure}[htb]
\centering
\includegraphics[width=0.8\textwidth,trim=90 40 40 100]{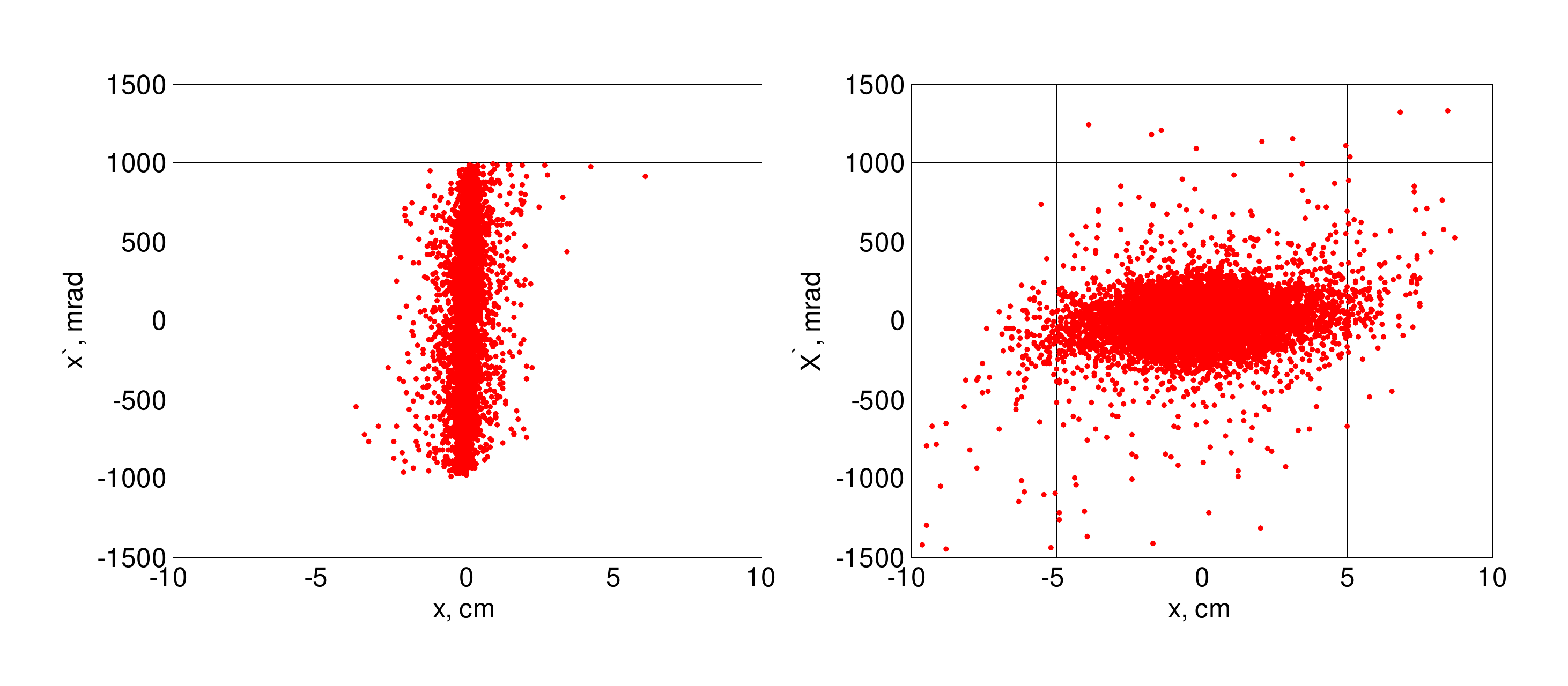}
\caption{Transverse phase portraits of the positron beam: left -- at AMD entry, right -- at exit.  \label{fig:12}}
\end{figure}

By tuning of the magnetic field distribution, we obtained required transformation of the transverse phase portrait of the positron beam, figure \ref{fig:12}.

\begin{figure}[htb]
\centering
\includegraphics[width=0.8\textwidth]{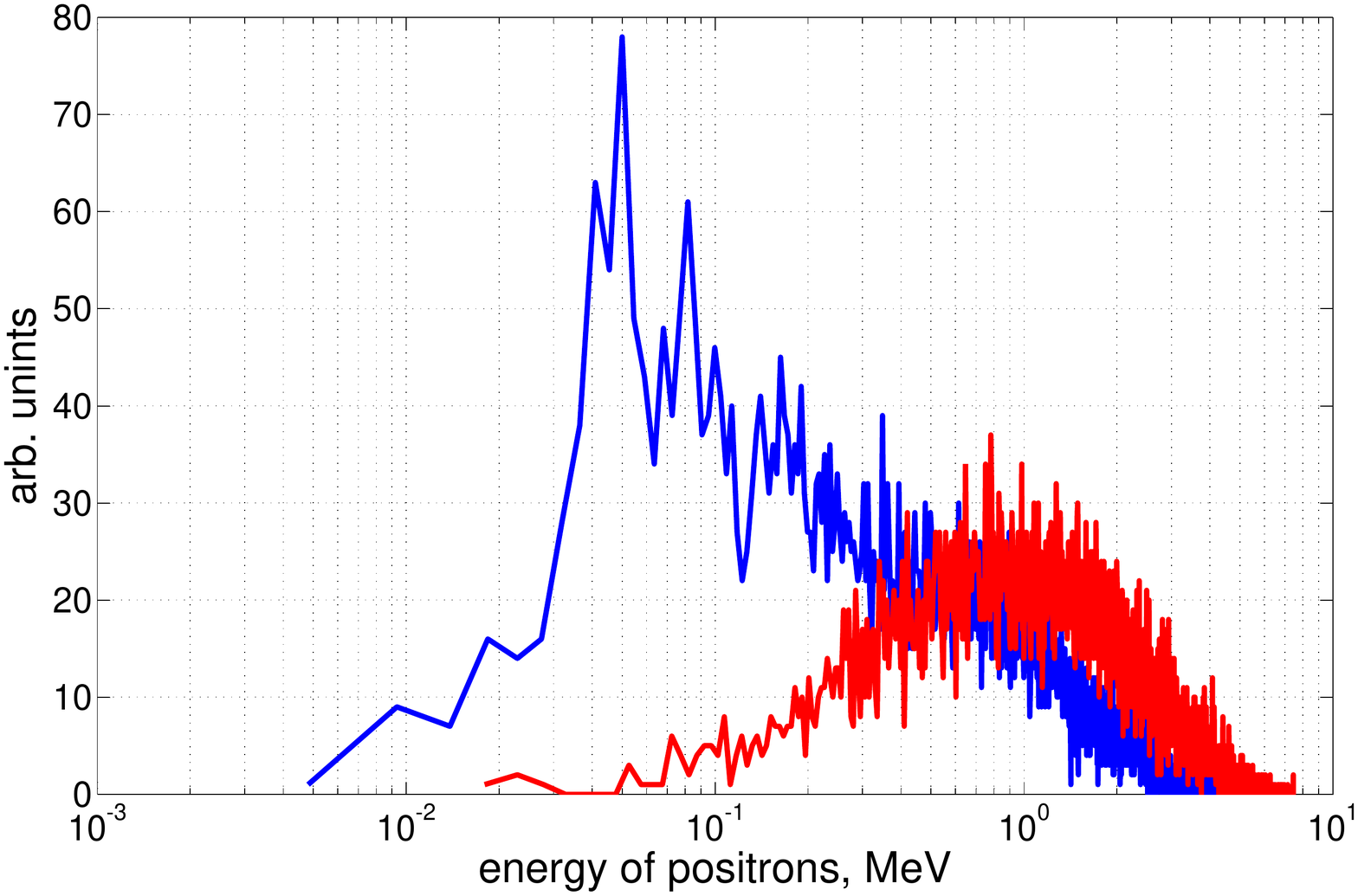}
\caption{Spectra of the positron beam: red curve at AMD entry, blue at exit.  \label{fig:13}}
\end{figure}

An optimal reduction of the positrons energy was obtained by variation of the RF cavity field amplitude and phase. The positron spectra before and after AMD are presented in figure \ref{fig:13} for the RF field strength 6\,MV/m.

As it can be seen from the figure, the cavity decrease energy of a vast fraction of the positron beam to the range $5\dots 200$\,keV. The maximum of positron spectrum is at 50\,keV, there is approximately 10\,\% of all positrons within the range $0\dots 200$\,keV.

\section{Summary and conclusion}
The process of the positrons production by the electron linacs has been considered theoretically. The problem of transporting positron beam from the electron-to-positron converter to a moderator that reduces energy of the positrons to desired level. Analysis on the adiabatic matching device (AMD) has been performed.

Dependence of the total positron yield on the energy of the initial accelerated electrons as well as on the thickness of the conversion target is obtained. As is was shown, there exists the optimal thickness of the target that maximizes the yield. This maximum is sufficiently broad that allows for variation the target thickness due to design considerations with no significant reduction of the yield.

Numerical simulations on the electron-to-positron conversion process have been performed for the electron energy from 9\,MeV to 90\,MeV. The conversion ratio and the spectrum of the positron beam, as well as optimal target thickness have been estimated, the results are in agreement with those obtained by other authors.

The simulation shows a significant number of the isotropically distributed annihilation photons, detecting of which allows to tune the positron source.

Qualitative analytical dependencies of the positron beam parameters at the system exit upon the amplitude and the decrease factor of the magnetic field in the AMD solenoid have been established. These dependencies have been used for optimizing the system.
Numerical simulations allow optimize the parameters of AMD for solenoid available in the laboratory.

Possible application of the subharmonic RF cavity for reduction of the energy of positrons is also estimated and validated by the simulations. As it was shown, this cavity can substantially decrease the positron energy and thus facilitate operation of the moderator.

The results obtained indicate a potential realization of the source of slow positrons on the linacs of NSC KIPT.

\subsection*{Acknowledgements}
The work is supported by the project X-6-1/2016 of National Academy of Science of Ukraine.


\begin{thebibliography}{22}
\providecommand{\natexlab}[1]{#1}
\providecommand{\url}[1]{\texttt{#1}}
\expandafter\ifx\csname urlstyle\endcsname\relax
  \providecommand{\doi}[1]{doi: #1}\else
  \providecommand{\doi}{doi: \begingroup \urlstyle{rm}\Url}\fi

\bibitem[Siegel(1980)]{siegel80}
R.W. Siegel.
\newblock Positron annihilation spectroscopy.
\newblock \emph{Annual Review of material science}, 10:\penalty0 393--425,
  1980.

\bibitem[Singh(2016)]{singh16}
Aditya~Narayan Singh.
\newblock Positron annihilation spectroscopy in tomorrow's material defect
  studies.
\newblock \emph{Applied Spectroscopy Reviews}, 51:\penalty0 1--48, 2016.

\bibitem[S.~Golde(2012)]{golde12}
B.~Vlaholic S.~Golde.
\newblock Review of low-energy positron beam facilities.
\newblock In \emph{Proc. of IPAC 12}, pages 1464--1466, 2012.

\bibitem[Wada et~al.(2016)Wada, Hyodo, Kasuge, et~al.]{wada13}
R.~Wada, T.~Hyodo, T.~Kasuge, et~al.
\newblock New experimental station at {KEK}. {S}low positron facility.
\newblock \emph{Journal of Physics: Conference Series}, 444:\penalty0 012082,
  2016.

\bibitem[O'Rourke et~al.(2013)]{rourke13}
B.~E. O'Rourke et~al.
\newblock Recent developments and future plans for the accelerator based slow
  positron facilities at {AIST}.
\newblock \emph{Materials Science Forum}, 733:\penalty0 285--290, 2013.

\bibitem[Chemerisov and Jonah(2009)]{chemerisov09}
S.~Chemerisov and C.~D. Jonah.
\newblock Developments of the positron facility at the argonne national
  laboratory's 20 {MeV} linac.
\newblock \emph{Materials Science Forum}, 607:\penalty0 243--247, 2009.

\bibitem[Artemov(1984)]{artemov84}
V.~I. Artemov.
\newblock \emph{Methods of production and forming of the positron beams in the
  linear accelerators}.
\newblock TsNIIatominform, Moscow, 1984.
\newblock (in Russian).

\bibitem[Akhiezer and Shulga(1993)]{shulga93}
A.I. Akhiezer and N.F. Shulga.
\newblock \emph{Electrodynamics of High Energies in Substance}.
\newblock Nauka, 1993.

\bibitem[Koch and Motz(1959)]{koch59}
H.~W. Koch and J.~W. Motz.
\newblock Bremsstrahlung cross-section formulas and related data.
\newblock \emph{Rev. Mod. Phys.}, 31:\penalty0 920, 1959.

\bibitem[Roy and Reed(1968)]{roy68}
R.~R. Roy and Robert~D. Reed.
\newblock \emph{Interactions of photons and leptons with matter}.
\newblock Academic Press, New York and London, 1968.

\bibitem[of~Standards and (NIST)(2016)]{nist}
National~Institute of~Standards and Technology (NIST).
\newblock {XCOM} code.
\newblock In \emph{\url{http://physics.nist.gov/cgi-bin/Xcom/xcom3_1}}. 2016.

\bibitem[toolkit for the simulation of the passage of particles~through
  matter(2007)]{geant4}
A~toolkit for the simulation of the passage of particles~through matter.
\newblock {Geant4} code.
\newblock In \emph{\url{https://cern.ch/geant4}}. 2007.

\bibitem[Dovbnya et~al.(2014)Dovbnya, Ayzatsky, Boriskin, et~al.]{dovbnya14}
A.N. Dovbnya, M.I. Ayzatsky, V.N. Boriskin, et~al.
\newblock State and prospects of the linac of the nuclear-physics complex with
  energy up to 100 {MeV}.
\newblock In \emph{Problems of Atomic Science and Technology, ser. ``Nuclear
  physics investigation''}, volume 3 (91), pages 60--63. 2014.

\bibitem[O'Rourke et~al.(2011{\natexlab{a}})O'Rourke, Hayashizaki, Kinomura,
  Kuroda, Minehara, Ohdaira, Oshima, and Suzuki]{rourke11}
B.~E. O'Rourke, N.~Hayashizaki, A.~Kinomura, R.~Kuroda, E.~Minehara,
  T.~Ohdaira, N.~Oshima, and R.~Suzuki.
\newblock Simulations of slow positron production using a low energy electron
  accelerator.
\newblock In \emph{\url{arXiv://physics/1102.1220v2}}. 2011{\natexlab{a}}.

\bibitem[Long et~al.(2007)Long, Chemerisov, Gai, Jonah, Liu, and Wang]{long07}
J.~Long, S.~Chemerisov, W.~Gai, C.~D. Jonah, W.~Liu, and H.~Wang.
\newblock Study on highflux accelerator positron source.
\newblock In \emph{Proc. of IPAC 07}, pages 2921--2923, 2007.

\bibitem[Chehab et~al.(1983)Chehab, LeMeur, Mouton, and Renard]{chehab83}
R.~Chehab, G.~LeMeur, B.~Mouton, and M.~Renard.
\newblock Adiabatic matching device for the {O}rsay linear accelerator.
\newblock \emph{IEEE Trans NS}, NS-30(4):\penalty0 2850--2852, 1983.

\bibitem[Chehab(1994)]{chehab94}
Robert Chehab.
\newblock Positron sources.
\newblock In \emph{CAS, Fifth General Accelerator Course. CERN 94-01, vol II},
  pages 643--678. 1994.

\bibitem[Lawson(1977)]{lawson}
J.D. Lawson.
\newblock \emph{The Physics of Charged--Particle Beams}.
\newblock Claredon Press, Oxford, 1977.

\bibitem[Bakay and Stepanovsky(1981)]{bakay81e}
A.S. Bakay and Yu.P. Stepanovsky.
\newblock \emph{Adiabatic invariants}.
\newblock Naukova Dumka, Kiev, 1981.
\newblock in Russian.

\bibitem[Young and Billen(2003)]{parmela}
L.~M. Young and J.~M. Billen.
\newblock The particle tracking code {PARMELA}.
\newblock In \emph{PAC 2003, Portland, USA}, page 3521, 2003.

\bibitem[O'Rourke et~al.(2011{\natexlab{b}})]{rourke}
B.~E. O'Rourke et~al.
\newblock Simulations of slow positron production using a low energy electron
  accelerator.
\newblock In \emph{\url{arXiv://physics/1102.1220v2}}. 2011{\natexlab{b}}.

\bibitem[Billen and Young(1993)]{superfish}
J.~H. Billen and L.~M. Young.
\newblock {POISSON/SUPERFISH} on {PC} compartibles.
\newblock In \emph{PAC 1993, Washington, USA}, pages 790--792, 1993.

\end{thebibliography}
\end{document}